\documentclass[aps,pre,twocolumn, titlepage]{revtex4-2}
\usepackage{graphicx}% Include figure files
\usepackage{color}
\usepackage{dcolumn}% Align table columns on decimal point
\usepackage{amsmath}
\usepackage{amsfonts}% Bold Huge math
\usepackage{bm}% bold math
\usepackage{epstopdf}
\usepackage{tabularx} % tabularx
\usepackage{multirow, booktabs}
\begin{document}

\title{Subharmonic Shapiro steps in depinning dynamics of a 2D solid dusty plasma modulated by 1D nonlinear deformed periodic substrates}

\author{Zhaoye Wang$^{1}$}
\author{Nichen Yu$^{1}$}
\author{Ao Xu$^{1}$}
\author{Chen Liang$^{1}$}
\author{C.~Reichhardt$^{2}$}
\author{C.~J.~O.~Reichhardt$^{2}$}
\author{Yan Feng$^{1,}$}
\thanks{The author to whom correspondence may be addressed: fengyan@suda.edu.cn}
\affiliation{$^{1}$ Institute of Plasma Physics and Technology, Jiangsu Key Laboratory of Frontier Material Physics and Devices, School of Physical Science and Technology, Soochow University, Suzhou 215006, China\\
{$^{2}$ Theoretical Division, Los Alamos National Laboratory, Los Alamos, New Mexico 87545, USA}
    }

\date{\today}

\begin{abstract}
Langevin dynamical simulations are performed to investigate the depinning dynamics of a two-dimensional (2D) solid dusty plasma, which is modulated by one-dimensional (1D) nonlinear deformed periodic substrates, and also driven by the combination of the DC and AC forces. As the DC driving force increases gradually, pronounced subharmonic and harmonic Shapiro steps are discovered under the combined modulation of the deformed substrate and the external AC drive. These observed subharmonic and harmonic Shapiro steps are attributed to the dynamic mode locking. The data analysis results indicate that the nonlinear deformed substrate strongly influences these subharmonic Shapiro steps, which can be accurately diagnosed using the fraction of sixfold coordinated particles. Furthermore, the diagnostic of the kinetic temperature clearly indicates the difference between harmonic and subharmonic Shapiro steps, i.e., the particle motion is slightly less synchronized at subharmonic Shapiro steps, caused by the unstable locations of the deformed substrate.

\end{abstract}

\maketitle

\section{Introduction}
The collective behavior of interacting particles under substrate modulation has attracted considerable interest in various two-dimensional (2D) physical systems, including colloids~\cite{PTiernoPRL:2012}, vortices in type-II superconductors~\cite{AEKoshelevPRL:1994}, Wigner crystals~\cite{C.ReichhardtPRB:2022}, electron crystals on a liquid helium surface~\cite{PMonceauAP:2012}, and dusty plasmas~\cite{LIWPRE:2019}. These external modulated substrates, such as one-dimensional (1D) periodic~\cite{LIWPRE:2019}, 2D periodic~\cite{CBechingerPRL:2001}, 1D vibrational periodic~\cite{youfengweiPRE:2022} and asymmetric~\cite{AVArzolaPRL:2011}, definitely play a crucial role in the collective behaviors of these physical systems. When a direct current (DC) driving force is applied to substrate-modulated systems, a variety of interesting depinning dynamics are generated, such as the directional locking~\cite{CReichhardtPRE:2004}, superlubricity~\cite{DMandelliPRB:2015}, commensuration effects~\cite{CReichhardtJPhys:2012, CReichhardtPC:2012}, kinks/antikinks~\cite{TBohleinNM:2012}, continuous/discontinuous
phase transitions~\cite{CReichhardtRPP:2017}, and reentrant
ordering~\cite{CReichhardtNJP:2023}. When an additional alternating current (AC) driving force is also introduced, more interesting dynamics can also further excited, such as the ratchet effect~\cite{CReichhardtPRE:2006}, bidirectional flow~\cite{LIWPRR:2023}, and Shapiro steps~\cite{CReichhardtPRB:2015}. 

Dynamic mode locking, or synchronization, was first observed by Huygens in pendulum clocks~\cite{PBakPT:1986}, which occurs in coupled oscillators when their different frequencies become locked under specific parameter conditions~\cite{TklingerPS:1997}. In the
presence of external both DC and AC driving forces, the motion of particles is manipulated by two different frequencies~\cite{BHuPRE:2007}, which are due to the DC and AC driving forces, respectively. As a result, the coupling between these two frequencies would lead to the synchronized motion of particles, resulting in a series of steps in the plot of the overall drift velocity with the varying DC driving force, which are termed as the Shapiro steps~\cite{MPNJunipernc:2015}, as first proposed for Josephson tunneling  currents~\cite{ShapiroPRL：1963}. When the locking occurs at integer multiples of the oscillation frequency, they are referred to as the harmonic Shapiro steps~\cite{BHuPRE:2007, TekicPRE:2010, JTekicPRE:2019}. While for the locking at non-integer rational multiples of the oscillation frequency, they are termed as the subharmonic Shapiro steps~\cite{BHuPRE:2007, TekicPRE:2010, JTekicPRE:2019}. From the previous investigations of substrate-modulated Frenkel-Kontorova models~\cite{BHuPRE:2005, M.PeyrardPRB:1982, TekicPRE:2010, JTekicJAP:2013}, to generate substantial subharmonic Shapiro steps, the asymmetric periodic substrates~\cite{BHuPRE:2005, TekicPRE:2010} or deformed periodic substrates~\cite{M.PeyrardPRB:1982, JTekicJAP:2013} are generally applied~\cite{BHuPRE:2005, TekicPRE:2010, JTekicJAP:2013}. Meanwhile, subharmonic Shapiro steps have also been studied in other physical systems, including the 2D Josephson arrays~\cite{SPBenzPRL:1990}, sliding charge-density waves~\cite{SEBrownPRL:1984}, and colloids~\cite{SVPTicco:2016}.

Dusty plasma, also referred to as complex plasma, is an excellent experimental model system consisting of dust particles, free electrons, ions, and neutral gas atoms~\cite{JBeckersPOP:2023, H.M.ThomasNature:1996, BonitzPRL:2006, NosenkoPRL:2004, HKahlertPRL:2012, CDuPRL:2019, LIscience:1996, AMelzerPRE:1996, JHChuPRL:1994, HThomasPRL:1994, RLMerlinoPhysT:2004, VEFortovPhysR:2005, GEMorfillRevModPhys:2009, ApielPlasmaPhysics:2010, MbonitzRRP:2010, LiangcPRR:2023, YFengPRL1:2008, EThomasPhysPlasmas:2004, WYUPRE:2022, FWiebenPRL:2019, YHePRL:2020}. Under the typical laboratory conditions, micron-sized dust particles are highly charged negatively to $\sim -10^{4}e$ within microseconds to the steady state~\cite{YFengPRE:2011}. The interaction between dust particles can be well described by the Yukawa repulsion~\cite{KonopkaPRL:2000} due to the shielding effects from free electrons and ions in plasma. Due to their high charges, these dust particles are strongly coupled, i.e., the interparticle electric potential energy is well beyond the averaged kinetic energy, exhibiting the typical properties of solids~\cite{YFengPRL1:2008} and liquids~\cite{EThomasPOP:2004}. These highly charged dust particles can be suspended and confined in the plasma sheath, so that they self-organize into a single layer, just the so-called 2D dusty plasma. Since they are immersed in plasma, dust particles experience a weak frictional gas drag as they move~\cite{B.Liu:2003}, with a typical damping rate of $\nu$ $\sim$ $1$ ${\rm s}^{-1}$~\cite{YFengPRL1:2008}. The motion of dust particles within this 2D plane can be precisely captured using video imaging and analyzed through particle tracking, so that various fundamental physical processes such as phase transitions~\cite{FengPRE:2008}, diffusion~\cite{BLIUPRL:2008}, and shock propagation~\cite{DSamsonovPRL:2004} can be investigated at the individual particle level.

Recently, various collective behaviors of 2D dusty plasma modulated by periodic substrates have been extensively investigated using computer simulations~\cite{LIWPRE:2019, XuaoPRR:2025, zhuwenqiPRE:2022, huangyuPRE2:2022, L.GuPRE:2020, zhuwenqiPOP:2023, ZWangPRE:2025}. While the substrate-modulated 2D dusty plasma is driven by the DC force with the increasing amplitude, the pinned, disordered plastic flow, and moving ordered states are observed~\cite{LIWPRE:2019}. Using different driving forces or different substrates, other interesting dynamics are also observed, such as the directional locking~\cite{zhuwenqiPRE:2022}, superlubricity~\cite{huangyuPRE2:2022}, commensuration effects~\cite{zhuwenqiPOP:2023}, the cyclic phase transition~\cite{XuaoPRR:2025}, bidirectional flow~\cite{LIWPRR:2023}, and the harmonic Shapiro steps~\cite{ZWangPRE:2025}. In Ref.~\cite{ZWangPRE:2025}, while the applied uniform DC driving force gradually increases, four harmonic Shapiro steps of the overall drift velocity are clearly observed in the substrate-modulated 2D solid Yukawa system. However, from our literature search, we have not found any previous reports of the subharmonic Shapiro steps in dusty plasmas, and the corresponding mechanism remains unclear, as we study here.

This paper is organized as follows. In Sec.~\ref{sec2}, we briefly describe our simulation method to study the behavior of 2D solid dusty plasma, which is modulated by the deformed substrates and also driven by the DC and AC forces. In Sec.~\ref{sec3}, we present our discovered subharmonic Shapiro steps in the studied 2D solid dusty plasma. We also perform systematic investigations of the corresponding structures and dynamics, so that our understanding of the corresponding mechnaism is provided. Finally, a summary of our findings is provided in Sec.~\ref{sec4}.

\section{Simulation methods}\label{sec2}
We follow the tradition~\cite{GJKalmanPRL:2004, H.OhtaPP:2000, K.Y.Sanbonmatsu:2001} to characterize the studied 2D solid dusty plasma using two dimensionless parameters, namely the coupling parameter $\Gamma$ and the screening parameter $\kappa$. They are defined as $\Gamma = Q^2 / 4 \pi \epsilon_{0} a k_{\rm B} T $~\cite{GJKalmanPRL:2004} and $\kappa = a / \lambda_{\rm D}$~\cite{GJKalmanPRL:2004}, respectively. Here, $Q$ is the charge on each particle, $T$ is the averaged kinetic temperature of dust particles, $a = (n \pi)^{-1/2} $ is the Wigner-Seitz radius with the 2D areal number density $n$, and $\lambda _{\rm D}$ is the Debye screening length. The length is normalized by using either the Wigner-Seitz radius $a$ or the lattice constant $b$. Note, for our studied 2D defect-free triangular lattice, $b = 1.90463a$~\cite{GJKalmanPRL:2004}.

Langevin dynamical simulations are performed to investigate subharmonic Shapiro steps in the substrate-modulated 2D solid dusty plasma driven by DC and AC forces in the $x$ direction. The equation of motion for each dust particle $i$ in our simulation is
\begin{equation}\label{EQyundong}
	m\mathbf{\ddot{r}}_i =- \sum \nabla \phi_{i j} - \nu m \mathbf{\dot{r}}_i + \mathbf{\zeta}_{i}(t) +\mathbf{F_{\rm s}} + \mathbf{F_{\rm d}} + \mathbf{F_{\rm a}}.
\end{equation}
Here, the first term on the right-hand-side (RHS) of Eq.~\eqref{EQyundong} comes from the binary Yukawa repulsion  $\phi_{ij}$ = $Q^2{\rm exp}(-r_{ij}/\lambda_{\rm D})/4\pi \epsilon_{\rm 0} r _{ij}$, where $r_{ij}$ is the distance between two particles~\cite{KonopkaPRL:2000}. The second term $-\nu m \boldsymbol{\dot{\mathbf{r}}}_i$ on the RHS is the frictional gas drag experienced by the moving particle in the rarefied gas or plasma environment~\cite{B.Liu:2003}. The third term $\zeta_{i}(t)$ represents the Langevin random kicks, which is assumed to have a Gaussian distribution with zero mean
value coming from the driven-dissipation theorem~\cite{FengPRE:2008}. The latter three terms of $\mathbf{F_{\rm s}}$, $\mathbf{F_{\rm d}}$, and $\mathbf{F_{\rm a}}$ come from the modulated substrate, external DC and AC forces, respectively, all in units of $F_{\rm 0} = Q^2/4 \pi \epsilon_{\rm 0} a^2$, as we explain in details next.

%Therefore, in our current investigation, following Ref.~\cite{M.PeyrardPRB:1982}, we 
To search for subharmonic Shapiro steps in dusty plasmas, we follow Refs.~\cite{BHuPRE:2005, M.PeyrardPRB:1982} to apply a 1D nonlinear deformed periodic substrate in our simulated 2D solid dusty plasma. In the previous studies~\cite{BHuPRE:2005, M.PeyrardPRB:1982}, it is found that the deformable substrates are able to induce subharmonic Shapiro steps. The expression of our applied substrate is specified as  
\begin{equation}\label{EQsubstrate}
      U(x) = \frac{U_{\rm 0}}{(2\pi)^2}(1-r)^2\frac{\left[1-\cos(2\pi x/w)\right]}{1+r^2+2r\cos\left(2 \pi x /w\right)},
\end{equation}
where $U_{\rm 0}$ and $w$ correspond to the depth and width of the substrate, in units of $E_{\rm 0} = Q^2 / 4 \pi \epsilon_{\rm 0} a$ and $b$, respectively. Here, $r$ is the shape parameter varying from $-1$ to $1$. In our current investigation, we specify $ U_{\rm 0} = 0.6 E_{\rm 0} $ and $ w = b $ in Eq.~\eqref{EQsubstrate}, with the shape parameter $r$ of $0$, $0.3$, $0.5$, and $0.7$, respectively. 

\begin{figure}
	\centering
	\includegraphics{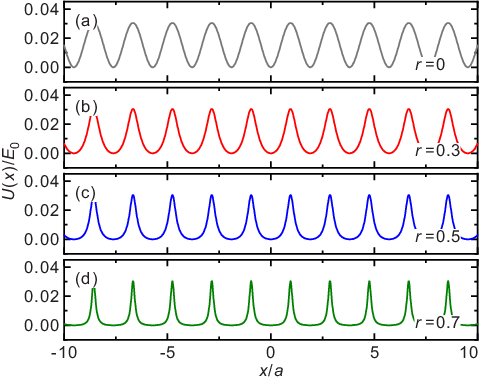}
	\caption{\label{fig1} Profiles of the nonlinear deformed substrate potential of Eq.~\eqref{EQsubstrate} for different values of the shape parameter $r = 0$ (a), $r = 0.3$ (b), $r = 0.5$ (c) and $r = 0.7$ (d). Clearly, as the shape parameter $r$ increases from $0$ to $0.7$, the substrate potential gradually deviates from the standard sinusoidal form, i.e., the peaks of the substrate potential become sharper. Note, the depth and width of the substrate potential are specified as constants of $U_{\rm 0} = 0.6 E_{\rm 0}$ and $w = b$, respectively.
    }
\end{figure}

To better illustrate the shape of applied substrate, we plot its potential profile in Fig.~\ref{fig1}. As shown in Fig.~\ref{fig1}(a), when $r = 0$, the substrate $U(x)$ exhibits the standard sinusoidal form. However, when $r$ increases gradually to $0.7$, the peaks of the substrate profile become sharper while the valleys are much flatter, resulting in the deformed substrates, as shown in Figs.~\ref{fig1}(b)-\ref{fig1}(d). Clearly, the force $\mathbf{F_{\rm s}}$ from the applied substrate can be analytically derived by combining $-\nabla U(x)$ and Eq.~\eqref{EQsubstrate}.

The driving DC force $\mathbf{F_{\rm d}}$ is applied along the $x$ direction with its magnitude increasing gradually from $0$ to $0.095 F_{\rm 0}$ with the interval of $10^{-3} F_{\rm 0}$ in our simulations. The applied AC driving force is $\mathbf{F_{\rm a}} = A\cos(2\pi f_{\rm s}t)\mathbf{\hat{x}}$, where $A$ and $f_{\rm s}$ are the amplitude and frequency, respectively. In our current investigation, we choose the constant values of $A = 2.5 F_{\rm 0}$ and $f_{\rm s} = 0.2 \omega_{\rm pd}$, where the nominal dusty plasma $\omega _{\rm pd} = (Q^2/ 2 \pi \epsilon_{\rm 0} m a^{3} )^{1/2}$~\cite{GJKalmanPRL:2004}.

The remaining details of our simulations are listed below. The conditions of our studied 2D dusty plasma are specified as $\Gamma =1000$, $\kappa = 2$, corresponding to the typical solid state~\cite{PHartmannPRE:2005}. The frictional gas damping rate is set as $\nu$ = 0.027$\omega_{\rm pd}$, similar to the typical value in 2D dusty plasma experiments~\cite{YFengPRE:2011}. All simulated $N_{\rm p} = 1024$ particles are confined inside a rectangular box with the dimensions of 60.9$a$ $\times$ 52.8$a$ using the periodic boundary conditions. We make sure that this simulation box is designed to contain exactly $32$ complete potential wells, so that periodic boundary conditions are always satisfied. For each simulation run, after the steady state is achieved, we integrate $\geq 6 \times 10^6$ steps with the time step of 0.001$\omega_{\rm pd}^{-1}$ to obtain the positions and velocities of all particles for the data analysis reported here. Other simulation details are the same as Ref.~\cite{LIWPRE:2019}.

\section{Results and discussions}\label{sec3}

\subsection{Subharmonic Shapiro steps}\label{3A}

\begin{figure*}
	\centering
	\includegraphics{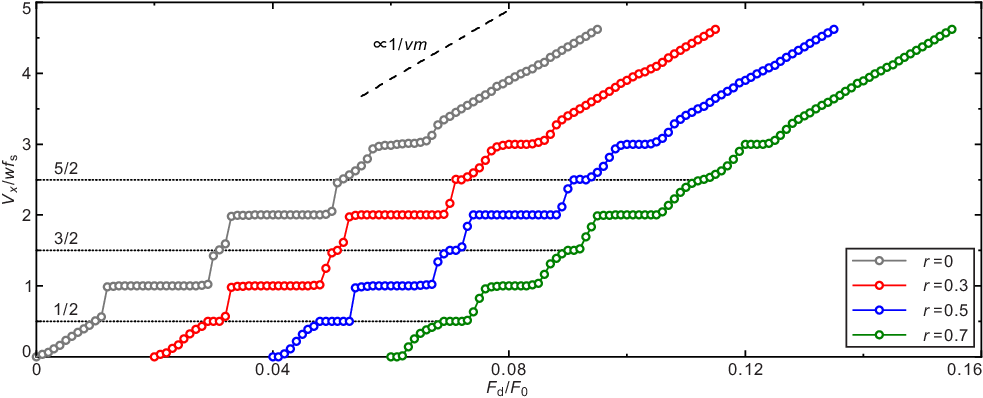}
	\caption{\label{fig2} Calculated collective drift velocity $V_{x}$ of a 2D Yukawa solid modulated by the 1D nonlinear deformed periodic substrate as the function of the applied uniform DC force $F_{\rm d}$, for different values of $r$. These $V_{x}$ results for different $r$ values are offset in $0.02F_{0}$ horizontally for clarity. Besides the harmonic Shapiro steps, three significant subharmonic Shapiro steps are discovered at $V_{x}/ wf_{\rm s} = 1/2$, $3/2$, and $5/2$ when $r = 0.3$ and 0.5, as the three dashed lines shown here. The three typical states of the depinning dynamics are also clearly exhibited here. Clearly, as $r$ increases, the critical depinning force also gradually increases. When $V_x/ wf_{\rm s} \ge 3.2 $ corresponding to $F_{\rm d}/F_{0} \ge 0.068$, the slope of each $V_{x}$ curve always approaches $1/\nu m$, indicating the moving ordered state there. In addition, when $r = 0.5$ or $0.7$, the stronger substrate deformation leads to the pinned and the plastic flow states. Note, the conditions of the applied AC driving force and substrate are always specified as $A = 2.5F_{\rm 0}$, $f_{\rm s} = 0.2 \omega_{\rm pd}$, $w = b$,  and $U_{\rm 0} = 0.6E_{\rm 0}$, respectively.
	} 
\end{figure*}

To investigate subharmonic Shapiro steps in the substrate-modulated 2D solid dusty plasma driven by the
DC and AC forces, we calculate the overall drift velocity along the $x$ direction using $V_{x} = N_{\rm p}^{-1} \left\langle \sum_{i=1}^{N_{\rm p}} {\mathbf{v}_{i} \cdot \hat{\mathbf{x}}} \right\rangle$. In this expression, $\mathbf{v}_{i}$ denotes the velocity of the particle $i$, while $\left\langle\right\rangle$ means the ensemble average for all particles. Our calculated results of $V_{x}$-$F_{\rm d}$ are presented in Fig.~\ref{fig2}, where the results for different values of $r$ are offset horizontally in a step of $0.02F_{\rm 0}$. Note, we confirm that the overall drift velocity in the $y$ direction $V_{y}$ is negligible due to the applied uniform external DC and AC forces only along the $x$ direction.

As the major result of this paper, we discover the occurrence of subharmonic Shapiro steps in the deformed-substrate-modulated 2D solid dusty plasma due to the combination of the DC and AC driving forces, as presented in Fig.~\ref{fig2}. From Fig.~\ref{fig2}, when $r=0$, $0.3$, $0.5$, and $0.7$, we observe three pronounced steps corresponding to $V_{x}/wf_{\rm s}=1$, $2$, and $3$, which are just the typical harmonic Shapiro steps~\cite{JTekicPRE:2019}. More importantly, when $r = 0.3$ and $0.5$, besides three harmonic Shapiro steps, we also discover that the overall drift velocity $V_{x}$ remains constant under the conditions of $V_{x}/wf_{\rm s}=1/2$, $3/2$, and $5/2$, just corresponding to the so-called subharmonic Shapiro steps~\cite{JTekicPRE:2019}, also termed the fractional Shapiro steps~\cite{S.MishraPRL:2025}. Similarly, when $r = 0.7$, we also find the subharmonic Shapiro steps at $V_{x}/wf_{\rm s}=1/2$ and $3/2$, however, there seems to be no visible subharmonic Shapiro step at $V_{x}/wf_{\rm s}=5/2$. Among these subharmonic Shapiro steps for various $r$ values, the step at $V_{x}/wf_{\rm s}=1/2$ is significantly wider than others, and the extreme widest one is approximately $ \geq 0.006 F_{0}$ when $r = 0.5$. 

Our observed Shapiro steps above can be expressed as
\begin{equation}\label{definition}
     V_{x} = \frac{p}{q}wf_{\rm s},
\end{equation}
where $p$ and $q$ are positive integers. As shown in Fig.~\ref{fig2}, we observe that the appearance of the harmonic Shapiro steps at $V_{x}/wf_{\rm s} = 1$, $2$, and $3$. While the subharmonic Shapiro steps occur at $V_{x}/wf_{\rm s} = 1/2$, $3/2$, and $5/2$, respectively. Clearly, when $q = 1$, Eq.~\eqref{definition} corresponds to harmonic Shapiro steps, while $q > 1$ represents subharmonic Shapiro steps. Here, we just term the steps at $V_{x}/wf_{\rm s} = 1/2$, $3/2$, and $5/2$ as the $1/2$, $3/2$, and $5/2$ subharmonic Shapiro steps, respectively. 

Besides the described harmonic and subharmonic Shapiro steps above, we also observe three typical states of the depinning dynamics~\cite{LIWPRE:2019} of the solid dusty plasma in Fig.~\ref{fig2}, which are the pinned, plastic flow, and moving ordered state. When $r = 0.5$ and 0.7, the pinned state are easily identified from $V_x = 0$ in Fig.~\ref{fig2}, and the critical depinning force increases with $r$. However, when $r = 0$ and $0.3$, the pinned state cannot be identified in Fig.~\ref{fig2}, maybe due to the relatively strong AC driving force, different from the previous studies~\cite{LIWPRE:2019, L.GuPRE:2020, zhuwenqiPOP:2023}. The lack of the pinned state and the subsequent plastic flow state resembles the superlubricity effect, similar to that observed in Ref.~\cite{huangyuPRE2:2022}. When $r$ is larger, i.e., $r = 0.5$ and 0.7, the solid dusty plasma is modulated by more deformed substrates, resulting in the pinned state and the subsequent plastic flow state. When the driving DC force is large enough, i.e., $F_{\rm d}/F_{0} \geq 0.068$, the drift velocity $V_{x}$ increases almost linearly with $F_{\rm d}$ with the constant slope of $1/m \nu$, which is just the characteristic of the moving ordered state~\cite{L.GuPRE:2020}.

\begin{figure}
	\centering
	\includegraphics{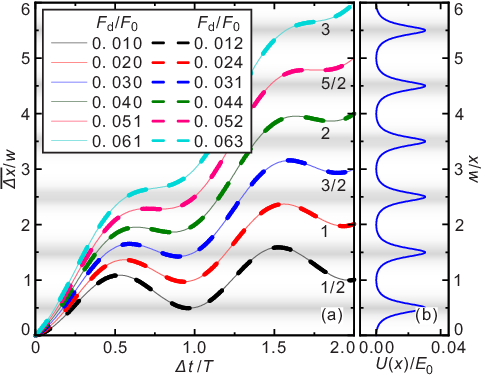}
	\caption{\label{fig3} Calculated particles’ average displacement $\overline{\Delta x}$ as the function of the time $\Delta t$ under the conditions of each Shapiro step at $r = 0.5$ (a), and the corresponding nonlinear deformed substrate of $r = 0.5$ (b). The shaded stripes represent the locations of the maxima of the substrate $U(x)$. Clearly, for each $p/q$ Shapiro step, the particle move over $p$ potential wells in $q$ periods. In addition, the oscillation of particle's average displacements indicate the pronounced back and forth motion of particles due to the applied AC driving force. 
	} 
\end{figure}

To intuitively illustrate the particle dynamics under the conditions of Shapiro steps, we present the average displacement $\overline{\Delta x}$ of the particles within two period under the condition of $r = 0.5$, as shown in Fig.~\ref{fig3}(a). Here, for each Shapiro step, we choose two different values of the driving force $F_{\rm d}$. Additionally, to better illustrate the particle motion, the corresponding deformed substrate of $r = 0.5$ is provided in Fig.~\ref{fig3}(b). Note, the shaded stripes in Figs.~\ref{fig3}(a) and \ref{fig3}(b) just indicate the locations of the substrate peaks.

Figure~\ref{fig3} clearly indicates the synchronized oscillatory drift of all particles under the combination of the DC and AC driving forces. As shown in Fig.~\ref{fig3}(a), we find that the curves of the average displacement $\overline{\Delta x}$ for two values of $F_d$ in the same step are exactly the same, indicating that all particles move in a synchronized manner under different driving forces in the same step. Furthermore, it can be also seen from the Fig.~\ref{fig3}(a) that all twelve curves exhibit oscillations. From our understanding, for our studied solid dusty plasma driven by both the DC and AC forces, the corresponding dynamics are characterized by the combination of two types of motion. These two types of motion are just the linear drift along the $x$ direction due to the DC force, and the back-and-forth oscillation caused by the AC force. As a result, from Fig.~\ref{fig3}, after one period of the AC driving force, the average drift displacement of particles is just the integer/half-integer times $w$ for the harmonic/subharmonic Shapiro steps.

\subsection{Mechanism of subharmonic Shapiro steps}\label{3B}

Let us first review the previous work of harmonic Shapiro steps observed in the 2D Yukawa solid~\cite{ZWangPRE:2025} due to dynamic mode locking. In Ref.~\cite{ZWangPRE:2025}, the 2D solid dusty plasma is modulated by a 1D vibrational periodic substrate and also driven by an uniform DC force. The vibrational periodic substrate~\cite{ZWangPRE:2025} provides the external AC modulation on the 2D solid dusty plasma. When the uniform DC driving force is increased gradually, four significant harmonic Shapiro steps are observed from the drift velocity as the function of the DC driving force, just corresponding to $V_{x}/wf_{\rm s}=1$, $2$, $3$, and $4$, respectively. In Ref.~\cite{ZWangPRE:2025}, these observed harmonic Shapiro steps are attributed to the dynamic mode locking, which is also termed as synchronization or phase locking. The dynamic mode locking describes a resonant interaction between the internal frequency and the externally applied oscillation frequency~\cite{TklingerPS:1997, Eott:2002}. Here, the internal frequency~\cite{MPNJunipernc:2015}, also known as the characteristic frequency~\cite{JTekicPRE:2019}, is caused by the motion of particles over a periodic substrate, since a uniform DC driving force is applied. While, the externally applied oscillation frequency directly comes from the AC driving~\cite{ZWangPRE:2025}. When the ratio of these two frequencies is an integer, harmonic Shapiro steps may appear~\cite{JodavicPRE:2015}. 

From our understanding, the observed subharmonic Shapiro steps above are probably caused by subharmonic mode locking~\cite{ZGShaoMPLB:2014, TekicPRE:2010}. In our current investigation, as the DC driving force gradually increases, the internal frequency of particles moving across potential wells also increases simultaneously. Meanwhile, the AC driving force $\mathbf{F_{\rm a}}$ provides an additional oscillatory frequency, which is fixed at $f_{\rm s} = 0.2 \omega_{\rm pd}$ in our simulation. When the ratio of the internal frequency to the oscillation frequency is a rational fraction, i.e., the locking occurs at fractional ratios, subharmonic Shapiro steps may occur. In these conditions, particles synchronously move across $p$ potential wells in $q$ periods. Thus, subharmonic Shapiro steps represent a resonant effect between the external AC driving force and the internal frequency of the depinning dynamics as all particles move over the potential wells of the deformed substrate. 

In fact, the modulation of the deformed substrate also has a significant effect on the occurrence of subharmonic Shapiro steps. As illustrated in Fig.~\ref{fig2}, under an increase of $0.001F_{\rm 0}$ in the driving force, the standard sinusoidal potential with $r=0$ exhibits only harmonic Shapiro steps, without any observable subharmonic Shapiro steps. However, when $r$ increases to $0.3$, the deformed substrate directly leads to clear subharmonic Shapiro steps without changing other conditions at all. We also notice that, in the optical potential experiments~\cite{APStikutsNC:2025} and simulations of other systems~\cite{ZGShaoMPLB:2014, BHuPRE:2005, JTekicJAP:2013}, the deformed substrates play a crucial role in the emergence of subharmonic Shapiro steps.

\begin{figure}
	\centering
	\includegraphics{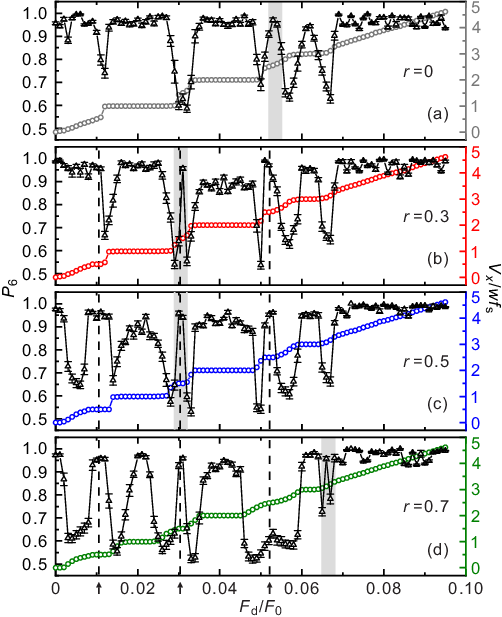}
	\caption{\label{fig4}  Calculated results of the fraction of sixfold coordinated particles $P_{\rm 6}$ and the corresponding drift velocity $V_{x}$ in Fig.~\ref{fig2} as the functions of the driving force $F_{\rm d}$ under the conditions of $r = 0$ (a), $r = 0.3$ (b), $r = 0.5$ (c) and $r = 0.7$ (d), respectively. Clearly, for the observed three $p/q$ subharmonic Shapiro steps indicated by the three dashed lines across panels (b)-(d) and harmonic Shapiro steps, as well as the pinned and moving ordered states, our calculated fraction of sixfold coordinated particles $P_{\rm 6}$ is always around $0.9$, indicating the highly ordered structure there. However, between these Shapiro steps, the values of $P_{\rm 6}$ are significantly lower, i.e., $<0.7$, corresponding to more disordered structures. The error bar of $P_{\rm 6}$ here comes from the standard deviation of the mean, obtained using $10^4$ data points of $P_{\rm 6}$ from $10^4$ frames. Note, more detailed analyzed results of the shaded regions are presented in Fig.~\ref{fig5}.
	} 
\end{figure}

\subsection{Structures around Shapiro steps}\label{3C}

)To characterize the structure of our studied system, we use the fraction of sixfold coordinated particles $P_{6}$~\cite{CReichhardtPRE:2005}. Here, $P_{6}$ is defined as $ P_{6} = N^{-1}_{\rm p}\left\langle \sum_{i=1}^N\delta(6-z_{i}) \right\rangle$ for each frame, where $z_{i}$ denotes the coordination number of particle $i$, obtained from the Voronoi construction~\cite{CReichhardtPRE:2005}. Note, $P_6 = 1$ for a perfect 2D triangular lattice, and this value decreases as the structure becomes more disordered, even approaching to zero in the limit of the highly disordered gas. In Fig.~\ref{fig4}, we plot the calculated $P_{6}$ and $V_{x}$ results as the functions of $F_{\rm d}$ under the conditions of $r = 0$, $0.3$, $0.5$, and $0.7$. While reporting the $P_{6}$ results in Fig.~\ref{fig4}, we calculate its standard deviation $\sigma$ using $10^{4}$ frames of particle arrangement, as the error bar shown there. Note, the dashed lines and the corresponding arrows in Figs.~\ref{fig4}(b)-\ref{fig4}(d) indicate the observed subharmonic Shapiro steps, while the shaded regions marked in Fig.~\ref{fig4} will be magnified in the latter presentation.

As illustrated in Fig.~\ref{fig4}, under the conditions of both subharmonic and harmonic Shapiro steps, the relatively high values of $P_{6}$ clearly indicate the highly ordered structures of our studied 2D solid dusty plasma. From our understanding, the motion of particles under these conditions are highly synchronized, reasonably leading to the highly ordered arrangements of particles, as reflected by the high values of $P_{6}$. Besides these Shapiro steps, the other three depining dynamics are also clearly exhibited, such as the moving ordered state when $F_{\rm d}/F_{0} \ge 0.068$ in Figs.~\ref{fig4}(a)-\ref{fig4}(d). However, the typical pinned and plastic flow states are only visible in Figs.~\ref{fig4}(c) and \ref{fig4}(d). For example, in Fig.~\ref{fig4}(c), the pinned state corresponds to the range of $0.000 \le F_{\rm d}/F_{0} \le 0.002$, where the related $P_{6}$ values are pretty high. While the ranges of $0.003 \le F_{\rm d}/F_{0} \le 0.007$ and $0.054 \le F_{\rm d}/F_{0} \le 0.058$ correspond to the plastic flow state in Fig.~\ref{fig4}(c), where the $P_{6}$ values are much lower. In Figs.~\ref{fig4}(a) and \ref{fig4}(b), we attribute the lack of pinned and plastic flow states to the superlubricity as in Ref.~\cite{huangyuPRE2:2022}, where the pinned state completely disappears so that particles exhibit the moving ordered state even though the driving force is extremely tiny.

\begin{figure}
	\centering
	\includegraphics{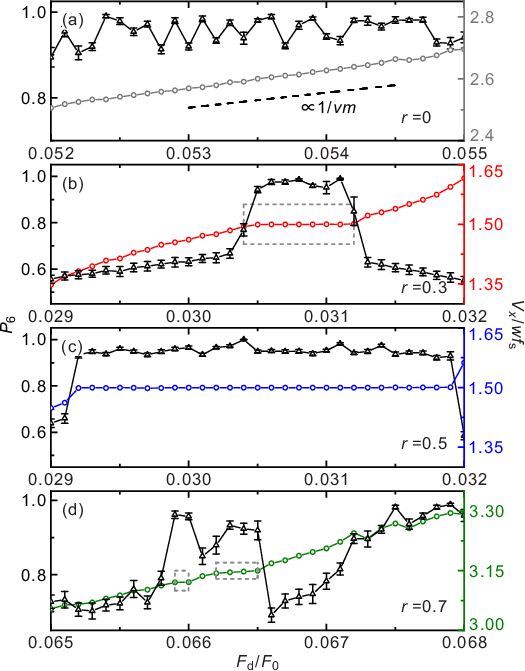}
	\caption{\label{fig5}    Magnified view of the $P_{\rm 6}$ and $V_{x}$ results corresponding to the shaded regions in Fig.~\ref{fig4}. Here, while performing Langevin dynamical simulations, we increase $F_{\rm d}$ with the interval of $10^{-4}F_{\rm 0}$, much smaller than $10^{-3}F_{\rm 0}$ in other figures. In panel (a), $V_{x}$ increases approximately linearly with a slope of $1/\nu m$ while $P_{\rm 6}$ remains at the high level, indicating the moving ordered state. In panels (b) and (c), the $3/2$ subharmonic Shapiro step is clearly observed, along with the consistent high level of $P_{\rm 6}$ there. In panel (d), two more subharmonic Shapiro steps are further found where the $P_{\rm 6}$ values are significantly higher. Note, in panels (b) and (d), we use dashed rectangles to mark the locations of subharmonic Shapiro steps from the unchanged values of $V_x$ and the higher values of $P_{\rm 6}$.
	}
\end{figure}

The results in Fig.~\ref{fig5} demonstrate that $P_{6}$ is able to sensitively diagnose subharmonic Shapiro steps. As shown in Figs.~\ref{fig5}(b) and \ref{fig5}(c), by using a much smaller increasing step of $F_{\rm d}/F_{0}=10^{-4}$, the $3/2$ subharmonic Shapiro step is much more obvious, while the $P_{6}$ results remain $>0.9$ throughout this subharmonic Shapiro step. Figure~\ref{fig5}(d) is mainly to confirm whether the only data point of $P_{6}$ jumping to a higher value in Fig.~\ref{fig4}(d) is reliable or not, since there is not any visible subharmonic Shapiro steps. However, in Fig.~\ref{fig5}(d), by confirming with the $P_{6}$ results, as $F_{\rm d}/F_{\rm 0}$ increases from $0.065$ to $0.068$, it seems that we find two more subharmonic Shapiro steps, as marked by the dashed rectangles, which might be the $25/8$ and $22/7$ steps, since $25/8$ and $22/7$ are both elements of the Farey sequence as predicted in Ref.~\cite{JodavicPRE:2015}. Note, although we only report the $1/2$, $3/2$, and $5/2$ subharmonic Shapiro steps in Figs.~\ref{fig2}-\ref{fig4}, as well as the two more subharmonic steps in Fig.~\ref{fig5}, by using smaller increasing step of $F_{\rm d}/F_{0}=10^{-4}$ and relying on the calculated $P_{6}$ results in all ranges more precisely, we believe that probably more new subharmonic Shapiro steps can be found in our studied systems.

\subsection{Dynamics of Shapiro steps}\label{4C}

\begin{figure}
	\centering
	\includegraphics{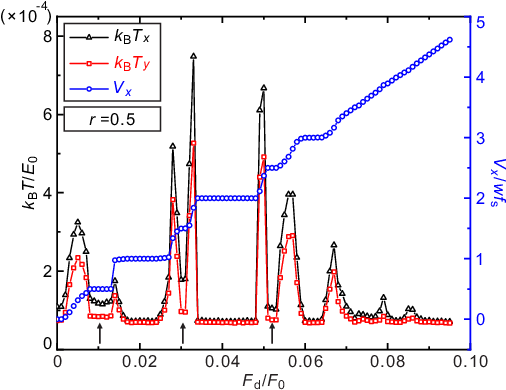}
	\caption{\label{fig6}  Calculated kinetic temperatures $k_{\rm B}T_{x}$ and $k_{\rm B}T_{y}$ from the particle motion in the $x$ and $y$ directions, as well as the corresponding drift velocity $V_{\rm x}$ for $r = 0.5$, as the functions of the DC driving force. Clearly, in the plastic flow state, $k_{\rm B}T_{x}$ and $k_{\rm B}T_{y}$ exhibit profound peaks, i.e., the values of $k_{\rm B}T_{x}$ and $k_{\rm B}T_{y}$ are much higher than the corresponding values in the Shapiro steps, the pinned and moving ordered states. At the corresponding subharmonic Shapiro steps, the values of $k_{\rm B}T_{x}$ and $k_{\rm B}T_{y}$ are significantly higher than those at the harmonic Shapiro steps, as indicated by the three arrows. Moreover, under the conditions of subharmonic Shapiro steps, $k_{\rm B}T_{x}$ is higher than $k_{\rm B}T_{y}$, which is different from the harmonic Shapiro steps, where $k_{\rm B}T_{x}$ and $k_{\rm B}T_{y}$ are almost the same low.
	}
\end{figure}

To further investigate the dynamics around the harmonic and subharmonic Shapiro steps, we also calculate the kinetic temperatures of $k_{\rm B}T_{x}$ and $k_{\rm B}T_{y}$ using the velocity fluctuations of all particles in the $x$ and $y$ directions, respectively. The kinetic temperature is calculated from the velocity fluctuations using $k_{\rm B}T = m\left\langle \sum_{i=0}^{N_{\rm p}}\left(\mathbf{v}_{i}-\overline{\mathbf{v}}\right)^2\right\rangle/2$, where the overall drift velocity $\overline{\mathbf{v}}$ is excluded. Here, $k_{\rm B}T_{x}$ and $k_{\rm B}T_{y}$ are presented in units of $E_{0}=Q^2 / 4 \pi \epsilon_{\rm 0} a$, as shown in Fig.~\ref{fig6},

Our calculated results of $k_{\rm B}T_{x}$ and $k_{\rm B}T_{y}$ in Fig.~\ref{fig6} clearly indicate the synchronized motion of particles under different conditions. From Fig.~\ref{fig6}, both $k_{\rm B}T_{x}$ and $k_{\rm B}T_{y}$ exhibit relatively low values at Shapiro steps, similar to the pinned and moving ordered states. In fact, for the subharmonic Shapiro steps, $k_{\rm B}T_{x}$ and $k_{\rm B}T_{y}$ are slightly higher than those at the harmonic Shapiro steps. The slightly higher values of the kinetic temperature for the subharmonic Shapiro steps clearly indicate that the motion of particles is less synchronized. Furthermore, as in Fig.~\ref{fig6}, $k_{\rm B}T_{x}$ and $k_{\rm B}T_{y}$ are almost the same for the harmonic Shapiro steps, due to the more synchronized motion of particles. However, for the less synchronized motion of subharmonic Shapiro steps, $k_{\rm B}T_{x}$ are substantially higher than $k_{\rm B}T_{y}$, since the DC and AC driving forces are both in the $x$ direction. In addition, around all termini of the harmonic and subharmonic Shapiro steps, $k_{\rm B}T_x$ and $k_{\rm B}T_y$ exhibit pronounced peaks, corresponding to the typically continuous transitions, as in Ref.~\cite{ZWangPRE:2025}.

From our understanding, as compared to the subharmonic Shapiro steps, the more synchronized particle motion observed on the harmonic Shapiro steps may be attributed to the effect of the dynamic mode locking mechanism. For the harmonic Shapiro step, particles move across $p$ potential wells within each single AC period. However, for the $p/q$ subharmonic Shapiro steps, particles tend to move from the valley to the peak of the specified substrate during the odd number of the period, as clearly shown in Fig.~\ref{fig3}. The motion of particles at peaks of the substrate, i.e., the typical unstable steady location, may result in higher velocity fluctuations. Thus, as compared to that for the harmonic Shapiro steps, the mode locking of the subharmonic Shapiro steps is definitely much more unstable, reasonably leading to the higher kinetic temperature observed in Fig.~\ref{fig6}.

\section{Summary}\label{sec4}
In summary, we investigate the depinning dynamics of a deformed-substrate-modulated 2D solid dusty plasma driven by the DC and AC forces using Langevin dynamical simulations. Besides the harmonic Shapiro steps as previous found in Ref.~\cite{ZWangPRE:2025}, we also discover the subharmonic Shapiro steps that occur when the ratio between the internal frequency and the oscillation frequency is a rational fraction. From our data analysis results, we attribute the discovered subharmonic Shapiro steps to the subharmonic mode locking. We also confirm that the nonlinear deformed substrate has an important effect on the formation of subharmonic Shapiro steps.

We also perform the systematic structural and dynamical analysis on these subharmonic Shapiro steps. These results clearly indicate that the fraction of sixfold coordinated particles $P_{6}$ remains consistently high under the conditions of both subharmonic and harmonic Shapiro steps, indicating the highly ordered structures under these conditions. Moreover, $P_{6}$ can be used as an efficient and sensitive diagnostic to detect more subharmonic Shapiro steps. The calculated kinetic temperature reveals the motion of particles becomes less synchronized at the subharmonic Shapiro steps than the harmonic Shapiro steps, probably due to the unstable locations of the specified nonlinear deformed substrates.

\subsection*{Acknowledgments}
The work was supported by the National Natural Science Foundation of China under Grant No. 12175159, the 1000 Youth Talents Plan, the Priority Academic Program Development of Jiangsu Higher Education Institutions, and the U. S. Department of Energy through the Los Alamos National Laboratory. Los Alamos National Laboratory is operated by Triad National Security, LLC, for the National Nuclear Security Administration of the U. S. Department of Energy (Contract No. 892333218NCA000001).

\bibliographystyle{apsrev4-1}

\end{document}